\documentclass[12pt]{article}
\usepackage{amsmath}
\usepackage{graphicx}
\usepackage{subfigure}

\newcommand{\ket}[1]{\ensuremath {|\: #1 \: \rangle}}
\newcommand{\bra}[1]{\ensuremath{\langle \: #1 \:|}}

\newcommand{\llrr}[1]{\ensuremath{\left( #1\right)}}
\newcommand{\llrrq}[1]{\ensuremath{\left[ #1\right]}}

\newcommand{\eref}[1]{(\ref{#1})}
\newcommand{\fref}[1]{figure \ref{#1}}

\begin{document}

\title{Dissipative dynamics of a spin system with three-body interaction}

\author{Diego de Falco and Dario Tamascelli\\
\small{Dipartimento di Scienze dell'Informazione, Universit\`a degli Studi di Milano}\\
\small{Via Comelico, 39/41, 20135 Milano- Italy}\\
\small{e-mail:defalco@dsi.unimi.it,tamascelli@dsi.unimi.it}
}
\date{}
\maketitle

\begin{abstract}
In this note we explicitly solve the Lindblad equation for a system of three spins with a three-body interaction, coupled to the environment by bath operators that inject or absorb spin carriers. We exemplify the properties of this solution in the context of a simple instance of Feynman's quantum computer in which a two-qubit program line is executed, applying the $\sqrt{NOT}$ primitive to a one-qubit register.
\\[5pt]
PACS: 03.67.Lx, 03.65.Yz
\end{abstract}

\section{Introduction \label{sec:intro}}
Much effort has been spent in recent years on the problem of quantum transport along a 1-dimensional spin chain with or without the presence of some interaction with an external environment\cite{bose03,dhar03,roy07,wichterich07,znidaric10}. There are, however, only few examples of exactly solvable open quantum systems\cite{breuer04,prosen08}. In this note we present the explicit solution of the Lindblad equation for a simple (the simplest, indeed) spin chain in which the interaction between two neighbouring sites is mediated by another spin located on the link that connects them. Such a three-body interaction is familiar in the context of lattice gauge theories \cite{wilson74} but we prefer to examine its role in the timing mechanism of Feynman's model of a quantum computer\cite{feyn86}.
\\In Feynman's machine, the application of computational primitives to the register is timed by a quantum clock: a single ``spin up'' travels along a spin chain and is coupled to the register in such a way that, when it moves from site $L$ to site $L+1$, the $L$-th operation is performed on the register. In an idealized mode of operation, one supposes that the timing excitation travels ballistically along the chain.\\
In this note we begin to explore the effect of excitations being injected/absorbed by the environment. Having in mind the Imry-Landauer \cite{landauer99} model of conductance, we model the environment through the bath operators suggested in section III.B of reference 4. Guided by the idea that computation is a nonequilibrium capability of a nonstationary system, we pay more attention, in this note, to short-time transient behaviour than to time-asymptotic properties.
\\[5pt]
The paper is organized as follows. In section II we introduce the model and review its ballistic behaviour. In section III we study the coupling  of the system to the environment via a Lindblad equation. Section IV is devoted to the discussion of examples. Conclusions and outlook are in section V. The spectrum of the Lindblad operator and a nonequilibrium stationary state (NESS) are studied in the appendix.
\section{The model: ballistic behaviour \label{sec:plain}}
\begin{figure}[h]
\centering
\includegraphics[width=7cm]{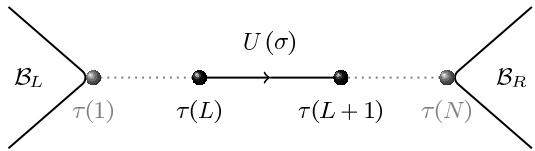}
\caption{The complete system: the endpoints of a spin chain of length $N$ interact with two reservoirs $\mathcal{B}_L$, $\mathcal{B}_R$. The interaction between sites $L$ and $L+1$ is mediated by an ancillary spin $\sigma$ through the unitary operator $U(\sigma)$.\label{fig:unit}}
\end{figure}
\begin{figure}[h]
\centering
\subfigure[]{\includegraphics[width=4.7cm]{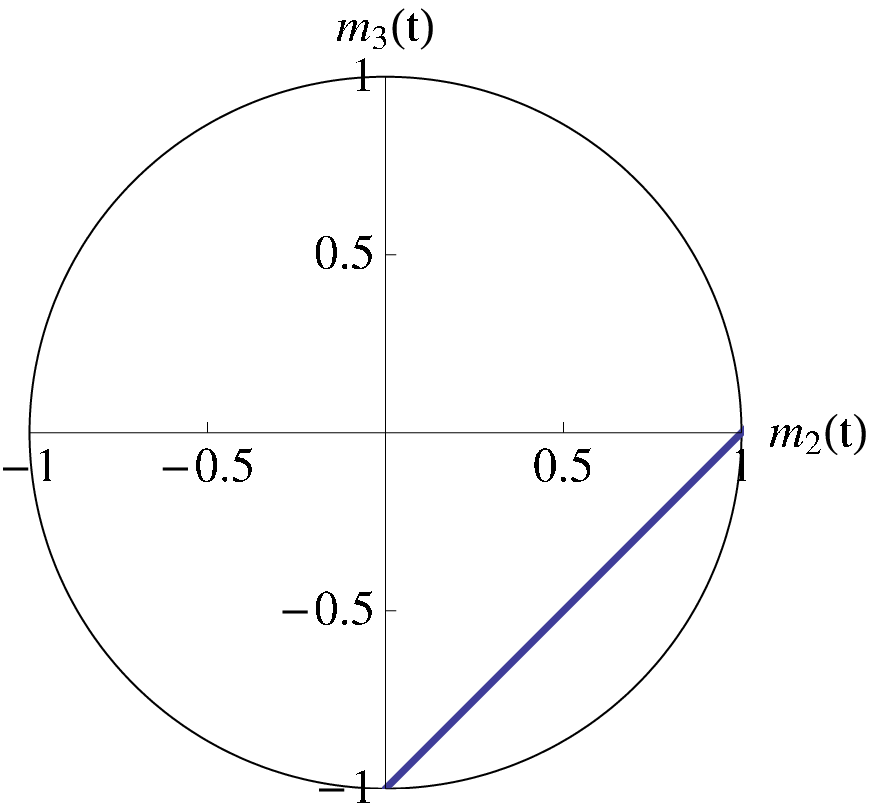}}
\subfigure[]{\includegraphics[width=5.2cm]{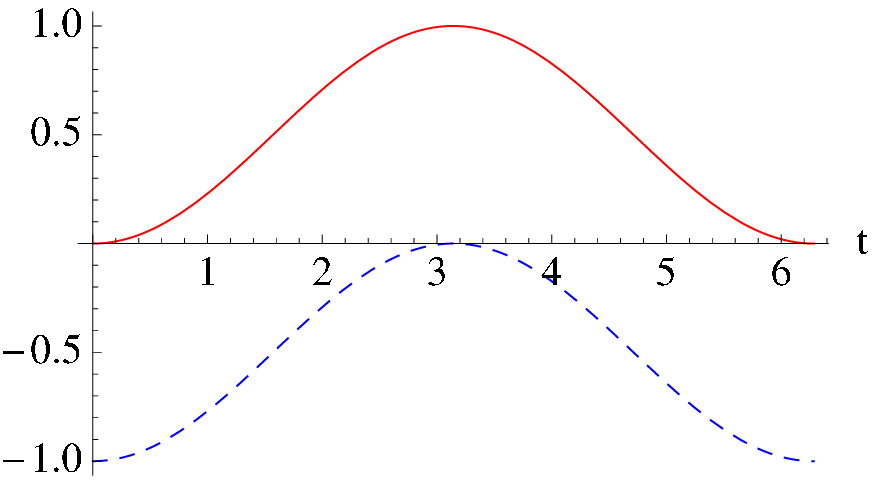}}\\
\subfigure[]{\includegraphics[width=5.2cm]{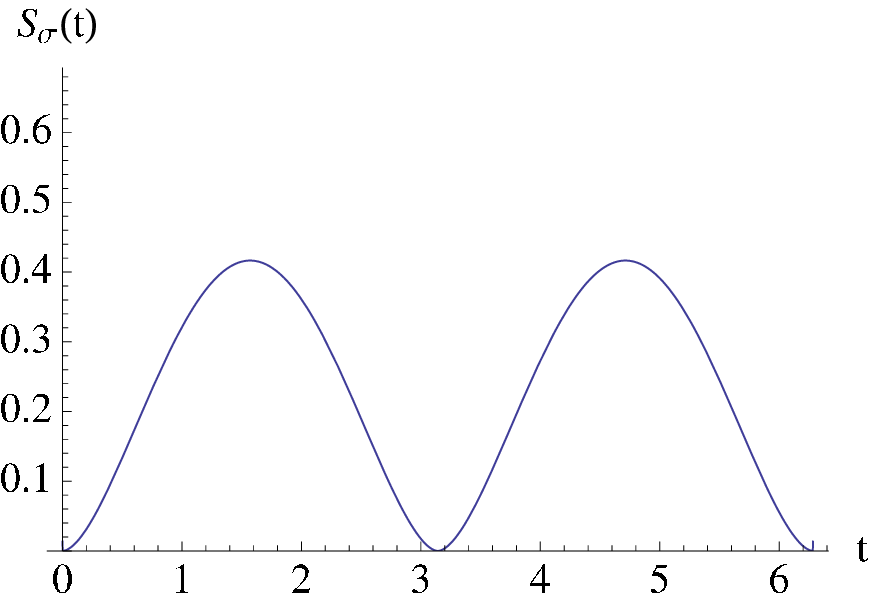}}
\subfigure[]{\includegraphics[width=5.2cm]{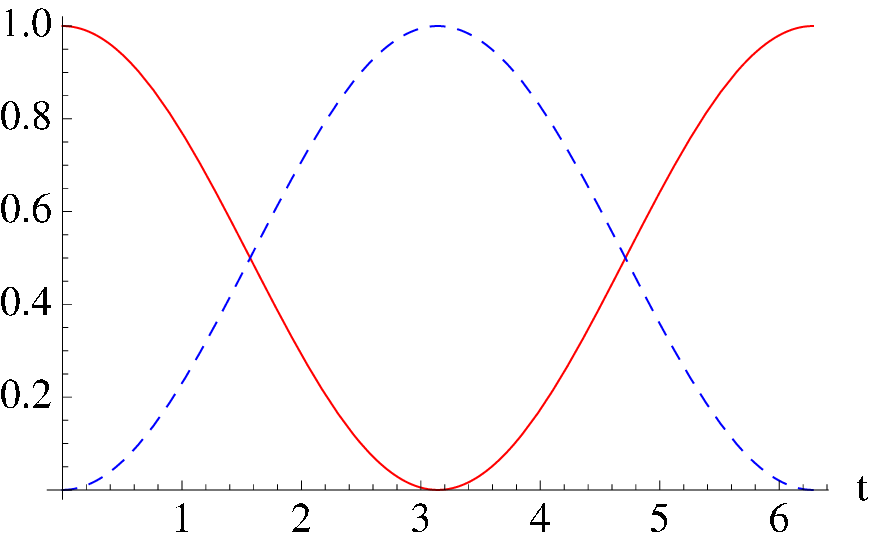}}
\caption{ Idealized mode of operation; $\vartheta=\pi/2$. (a) A parametric plot of $\llrr{m_2(t),m_3(t)}\ \mbox{for } 0 \leq t \leq 2 \pi$ shown inside the unit circle. (b) $m_2(t)$ (solid line) and $m_3(t)$ (dashed line) as functions of time. (c) The von Neumann entropy of the register as a function of time. (d) Solid  line: $n_3(L,t)$, dashed line: \mbox{$n_3(L+1,t)$}.\label{fig:figFeynnoint}}
\end{figure}

Figure \ref{fig:unit} represents the basic functional unit of Feynman's model of a quantum computer\cite{feyn86,peres85,cirac08,defa06b,nagaj10}.
The spin 1/2 systems located at the vertices $L$ and $L+1$ of a graph, that we indicate by $\tau(L)=\llrr{\tau_1(L),\tau_2(L),\tau_3(L)}$ and $\tau(L+1)=\llrr{\tau_1(L+1),\tau_2(L+1),\tau_3(L+1)}$, are part of the \emph{clocking mechanism} or \emph{cursor}.\\
We will make the simplification of neglecting all other cursor sites ($N=2$). This will be crucial in making our model explicitly solvable. The effect of a long clocking chain is, by the way, quite well understood \cite{defa06a,prosen08} and we wish to concentrate here on other effects.
\\In an \emph{idealized} mode of operation (to be reviewed below), the vertices $L$ and $L+1$ support an excitation that, when travelling along the oriented edge \llrr{L,L+1}, applies the unitary transformation $U(\sigma)$ to a \emph{register} two-level system $\sigma = \llrr{\sigma_1,\sigma_2,\sigma_3}$. The inverse transformation $U^{\dagger}(\sigma)$ is similarly associated with the oriented edge $\llrr{L+1,L}$.\\
We will suppose in this section that, setting, in our simple model, $L=1$ and $\tau_{\pm}(j) = \llrr{\tau_1(j) \pm i \tau_2(j)}$, for $j=L,L+1$, the above degrees of freedom are coupled by a Hamiltonian of the form:
\begin{align}{\label{eq:freeham}}
H &= -\frac{1}{2} U(\sigma)\  \tau_+(L+1)\tau_-(L)  -\frac{1}{2} U^\dagger(\sigma)\  
\tau_-(L+1) \tau_+(L).
\end{align}
We will take, by way of example, \mbox{$U(\sigma) = e^{-i \frac{\vartheta}{2} \sigma_1}$}, namely a rotation by an angle $\vartheta$ around the axis 1.\\
With this choice, the Hamiltonian $H$ is best described in a basis of simultaneous eigenvectors of \llrr{\sigma_1,\tau_3(L),\tau_3(L+1)}. We will list the vectors of this basis in lexicographic order (with ``1'' preceding ``-1'') as:
\begin{align*}
&\ket{e_1} = \ket{\sigma_1=1,\tau_3(L)=1,\tau_3(L+1)=1}\\
&\ket{e_2} = \ket{\sigma_1=1,\tau_3(L)=1,\tau_3(L+1)=-1}\\
&\ldots \\
&\ket{e_8} = \ket{\sigma_1=-1,\tau_3(L)=-1,\tau_3(L+1)=-1}.
\end{align*}
In this representation the only non vanishing matrix elements of $H$ are:
\begin{align*}
&\bra{e_2}H\ket{e_3} = \bra{e_7}H\ket{e_6} = -\frac{1}{2} e^{i \frac{\vartheta}{2}} = \overline{\bra{e_3}H\ket{e_2}} = \overline{\bra{e_6}H\ket{e_7}}.
\end{align*}
The matrix $\left\| \bra{e_j} H \ket{e_k}\right \|_{j=1,\ldots,8 \atop k=1,\ldots,8} $ is easily diagonalized and exponentiated leading to a quite explicit expression of the time evolution operator $Z(t) = \exp(-itH)$. 
In particular, for any initial density matrix $\rho_{in}$ we can express the solution of the quantum Liouville equation $d \rho_t/dt = -i [H,\rho_t]$ as $\rho_t = Z(t) \rho_{in} Z^\dagger(t)$.
\\In this note, where not otherwise stated, we will always suppose that the state evolves from the following initial condition:
\begin{equation}{\label{eq:initialcond}}
\rho_0 = \frac{1}{2^3} (I-\sigma_3) (I+\tau_3(L))(I-\tau_3(L+1)). 
\end{equation}
This is the typical initial state of a Feynman machine: the excitation is initially located at the beginning of the spin chain, and the register $\sigma$ is in the state corresponding to the input ($\sigma_3=-1$ in our specific example).
\\Under this initial condition, the probability that \mbox{$\tau_3(L)=1$} is given by:
\[
n_3(L,t) = Tr\llrrq{\rho_t \frac{I+\tau_3(L)}{2}} = \cos\llrr{\frac{t}{2}}^2;
\]
the probability that $\tau_3(L+1)=1$ is given by:
\[
n_3(L+1,t) = Tr\llrrq{\rho_t \frac{I+\tau_3(L+1)}{2}} = \sin\llrr{\frac{t}{2}}^2.
\]
While the ``spin up'' excitation bounces back and forth between the two \emph{cursor} sites $L$ and $L+1$, the \emph{register} spin $\sigma$ is acted upon by the operators $U$ and $U^\dagger$, so that its Bloch vector evolves as:
\begin{align*}
&m_1(t) = Tr\llrrq{\rho_t\ \sigma_1} = 0; \\
&m_2(t) = Tr\llrrq{\rho_t\ \sigma_2} = \sin\llrr{t/2}^2 \sin(\vartheta); \\
&m_3(t) = Tr\llrrq{\rho_t\ \sigma_3} = \sin\llrr{t/2}^2 (1-\cos(\vartheta))-1.
\end{align*}
We adopt, in most of the examples that follow, the value $\vartheta = \pi/2$. We refer to the primitive $U(\pi/2)$ as to $\sqrt{NOT}$, because by applying it twice the initial $\sigma_3=-1$ state gets flipped into $\sigma_3=+1$.\\
For the sake of comparison with the behaviour studied in the following sections, it is interesting to look at the Bloch diagram of the register spin, shown in \fref{fig:figFeynnoint}(a). The motion of the system is periodic: at times $t=(2k-1)\pi,\ k=1,2,\ldots$, the computation $\sqrt{NOT}$ is completed and the state of the register is on the Bloch surface, i.e. its the von Neumann entropy is zero. At times $t=2k\pi$ the computation turns out to be undone, the state is brought back to the surface of the Bloch sphere, the clocking excitation is brought back to its initial position.
\\Of particular interest for the considerations that follow is the graph (\fref{fig:figFeynnoint}(c)) of the von Neumann entropy of the register:
\[
S^{reg}(t) = -\frac{1+r(t)}{2}\ln\llrr{\frac{1+r(t)}{2}} - \frac{1-r(t)}{2}\ln\llrr{\frac{1-r(t)}{2}}
\]
where $r(t)= \sqrt{m_2(t)^2 + m_3(t)^2}$. Since the system is a closed bipartite quantum system, its overall entropy is zero, and the two subsystems, register and cursor, are isoentropic. The fact that there are times at which the intended computation (the rotation by $\pi/2$ of a spin initially pointing ``down'') is completed \emph{with certainty} is related to the possibility of bringing, at such times, $S^{reg}(t)$ to its initial value 0. Any source of noise, therefore, might blur such a certainty.
\\As a final remark of this section we wish to give a hint to the development of the entanglement during the evolution under the Hamiltonian \eref{eq:freeham} of the initial condition \eref{eq:initialcond}. To this end we study the mean value of the chirality operator:
\[
\chi = \sigma \cdot \tau(1)\times\tau(2),
\]
whose interest as an entanglement witness is discussed in reference 16. It is easy to check that:
\[
Tr\llrr{\chi\ \rho_t}=2 \cos(\vartheta/2) \sin(t)
\]
so that it is only for $\vartheta=0$ that the state of the system attains the maximum value 2 of mean chirality compatible with bipartite entanglement.

\section{Lindblad equation \label{sec:open}}
The model outlined in the previous section is so simple that we can try to study the effect on it of  interaction with the environment.
\\Following references 4 and 5, we describe this interaction in the Markovian approximation and adopt a Lindblad equation approach, in which the following \emph{bath operators} act on the cursor spins:
\begin{align}{\label{eq:lindops}}
&L_1 = \sqrt{\epsilon (1-\mu)} \tau_+(L), &\qquad &L_2 = \sqrt{\epsilon (1+\mu)} \tau_-(L),\\
&L_3 = \sqrt{\epsilon (1+\mu)} \tau_+(L+1), &\qquad &L_4 = \sqrt{\epsilon (1-\mu)} \tau_-(L+1), \nonumber
\end{align}
where the coupling parameter $\epsilon$ and the asymmetry parameter $\mu$ satisfy the conditions $\epsilon >0$ and $-1 \leq \mu \leq 1$. The asymmetry parameter $\mu$ models a possible difference of the chemical potentials of $\mathcal{B}_L$ and $\mathcal{B}_R$. We refer the reader to the second section of reference 17 for a discussion of the range of validity of this approach and of the possible degree of control of the baths and of the bath-system interaction.\\
The density operator is supposed to evolve according to the equation:
\begin{equation}{\label{eq:lindblad}}
\frac{d\rho_t}{dt} = -i \llrrq{H,\rho_t}-\frac{1}{2} \sum_{j=1}^4\left \{ L_j^\dagger L_j,\rho_t \right \}+ \sum_{j=1}^4 L_j \rho_t L_j^\dagger.
\end{equation}
The main aim of this note is to show that the Cauchy problem posed by \eref{eq:initialcond}, \eref{eq:lindops} and \eref{eq:lindblad} can be given an explicit solution, parametrized by $\vartheta,\ \epsilon,\ \mu$. This is shown in full detail in the appendix. The full spectrum of the Lindblad matrix is, furthermore, provided there for the reader interested in initial conditions other than \eref{eq:initialcond}.
\\In this section we study some physical properties of \emph{this} solution.
\\We observe, first of all, that we can give a quite explicit expression for the joint probability distribution of $\tau_3(L)$ and $\tau_3(L+1)$; setting $\rho_t(j,k)= \bra{e_j}\rho_t \ket{e_k}$ and using the solution (see the appendix) of $System(+,+)$ we have:
\\[5pt]
\emph{i.} Probability that $\tau_3(L)=1 \wedge \tau_3(L+1)=1$ and probability that $\tau_3(L)=-1$ $\wedge\tau_3(L+1)=-1$:
\begin{align*}
& Tr\llrrq{\rho_t \llrr{\frac{I + \tau_3(L)}{2}} \llrr{ \frac{I+\tau_3(L+1)}{2}}} =\\
& =Tr\llrrq{\rho_t \llrr{\frac{I - \tau_3(L)}{2}} \llrr{ \frac{I-\tau_3(L+1)}{2}}}= \rho_t(1,1)+ \rho_t(5,5) =\\
& = \frac{1+ 4 \epsilon^2 \llrr{1-\mu^2}}{4\llrr{1+4 \epsilon^2}} - e^{-4 t \epsilon} \frac{1+4 \epsilon^2 \llrr{1+\mu}^2}{4 \llrr{1 + 4 \epsilon^2}}+ \\
& +e^{-2 t \epsilon} \frac{\epsilon \mu \llrr{2 \epsilon\llrr{1+\mu}\cos(t)+\sin(t)}}{1+4\epsilon^2}.
\end{align*}
We observe that these two probabilities vanish for $\epsilon=0$ (conservation of the number of clocking excitations in the Hamiltonian regime); for $\epsilon>0$ the bath operators can kill the single clocking excitation introduced in the initial condition \eref{eq:initialcond} or introduce a spurious one.
\\[5pt]\emph{ii.} Probability that $\tau_3(L)=-1 \wedge \tau_3(L+1)=1$:
\begin{align*}
 &Tr\llrrq{\rho_t \llrr{\frac{I - \tau_3(L)}{2}} \llrr{ \frac{I+\tau_3(L+1)}{2}}}=\rho_t(3,3)+\rho_t(7,7)= \\
 &=\frac{1+ 4 \epsilon^2 \llrr{1+\mu}^2}{4\llrr{1+4 \epsilon^2}}+ e^{-4 t \epsilon} \frac{1+4 \epsilon^2 \llrr{1+\mu}^2}{4 \llrr{1 + 4 \epsilon^2}}-e^{-2 t \epsilon} \frac{\llrr{1+4 \epsilon^2\llrr{1+\mu}^2}\cos(t)}{2\llrr{1+4\epsilon^2}}.
\end{align*}
\\[5pt]\emph{iii.} Probability that $\tau_3(L)=1 \wedge \tau_3(L+1)=-1$:
\begin{align*}
 &Tr\llrrq{\rho_t \llrr{\frac{I + \tau_3(L)}{2}} \llrr{ \frac{I-\tau_3(L+1)}{2}}}= \rho_t(2,2)+\rho_t(6,6)= \\
 &=\frac{1+ 4 \epsilon^2 \llrr{1-\mu}^2}{4\llrr{1+4 \epsilon^2}}+ e^{-4 t \epsilon} \frac{1+4 \epsilon^2 \llrr{1+\mu}^2}{4 \llrr{1 + 4 \epsilon^2}} +\\
&+ e^{-2 t \epsilon} \frac{\llrr{1+4 \epsilon^2 \llrr{1-\mu^2}}\cos(t)-4 \epsilon \mu \sin(t)}{2\llrr{1+4\epsilon^2}}.
\end{align*}
The fact that the above joint distribution of $\tau_3(L)$ and $\tau_3(L+1)$ does not depend on the parameter $\vartheta$ justifies Feynman's statement that ``\emph{it turns out that the propagation of the cursor up and down this program line is exactly the same as it would be if the operator [U] were not in the Hamiltonian}''\cite{feyn86}.\\
This statement needs however a qualification because of the $\vartheta$-dependence of some components of the solution of $System(+,+)$ and $System(-,-)$ exhibited in the appendix: there are, in fact, observables of the cursor (other than $\tau_3(L)$ and $\tau_3(L+1)$) whose distribution depends on the primitive being applied to the register. For instance, for the current operator, one has:
\begin{align*}
&Tr\llrrq{\rho_t \frac{i}{2} \llrr{\tau_-(L) \tau_+(L+1)-\tau_-(L+1)\tau_+(L)}} =\\
& = \frac{i}{2}\llrr{\rho_t(2,3) - \rho_t(3,2) +\rho_t(6,7)-\rho_t(7,6)} = \\
&= - \cos\llrr{\frac{\vartheta}{2}}\frac{4 \epsilon \mu}{4 \llrr{1+4 \epsilon^2}}+\\
& + \cos\llrr{\frac{\vartheta}{2}} \frac{2 e^{-2t\epsilon} \llrr{2 \epsilon \mu \cos(t)+\llrr{1+ 4 \epsilon^2 (1+\mu)}\sin(t)}}{4 \llrr{1+4 \epsilon^2}}.
\end{align*}
As to the study  of the Bloch diagram of the register, we observe that, setting 
\[
m(t) = 2 \sum_{j=1}^4 \rho_t(j,j+4),
\] 
it is:
\begin{align*}
&m_2(t) = Tr\llrrq{\rho_t \ \sigma_2} = Im\llrr{m(t)},\\
&m_3(t) = Tr\llrrq{\rho_t \ \sigma_3} = Re\llrr{m(t)}.
\end{align*}
Under the initial condition \eref{eq:initialcond} and under the assumption that the four eigenvalues $\omega_i$ defined in the appendix are distinct, we can give an explicit expression of the function $m(t)$ in terms of the solution of $System(+,-)$.
\\[5pt]
Having the analytic solution of the master equation allows us to make some considerations about the asymptotic behaviour of the system as well.\\
For example, the relaxation time of the register is determined by
\[
\frac{1}{max \{Re\llrr{\omega_i},i=1,\ldots,4\}}.
\] 
The relaxation time of the cursor, on the other side, is $(2\epsilon)^{-1}$, as it is possible to evince from the explicit expression of the matrix elements of $System(+,+)$.\\
Other aspects of the system quantitatively accessible through the analytic solution are the asymptotic entropies of the overall system and of its components. It is easy to check that, for the register, it is:
\[
\lim_{t \to \infty}S^{reg}(t) = \log(2).
\]
For the overall system it is:
\[
\lim_{t \to \infty} Tr\llrrq{- \rho_t \log\llrr{\rho_t}} = \log (8) - 4 \epsilon^2 \mu^2 + O(\epsilon^4).
\]
Setting $\rho_t^{cur} = Tr_\sigma(\rho_t)$, it is easy to check that the asymptotic entropy of the cursor 
\begin{align*}
 \lim_{t \to \infty}  Tr\llrrq{-\rho_t^{cur} \log \llrr{\rho_t^{cur}}} = \log (4)-2 \epsilon ^2 \mu ^2 (\cos (\theta )+1)+ O\llrr{\epsilon^3}
\end{align*}
depends on the \emph{three} parameters.\\
As a final remark of this section, we point out that the mean chirality is given by:
\[
Tr\llrr{\chi\ \rho_t}=8\ Im(\rho_t(2,7)).
\]

\section{Examples \label{sec:examples}}
\begin{figure}[h]
\centering
%
\subfigure[]{\includegraphics[width=4.7cm]{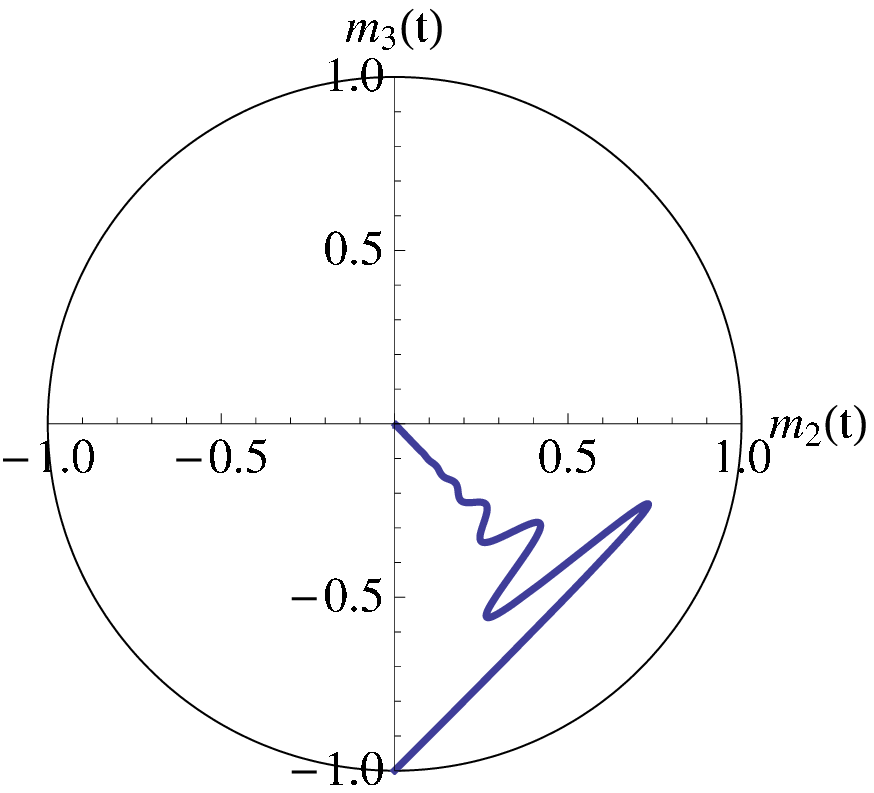}}
\subfigure[]{\includegraphics[width=5.2cm]{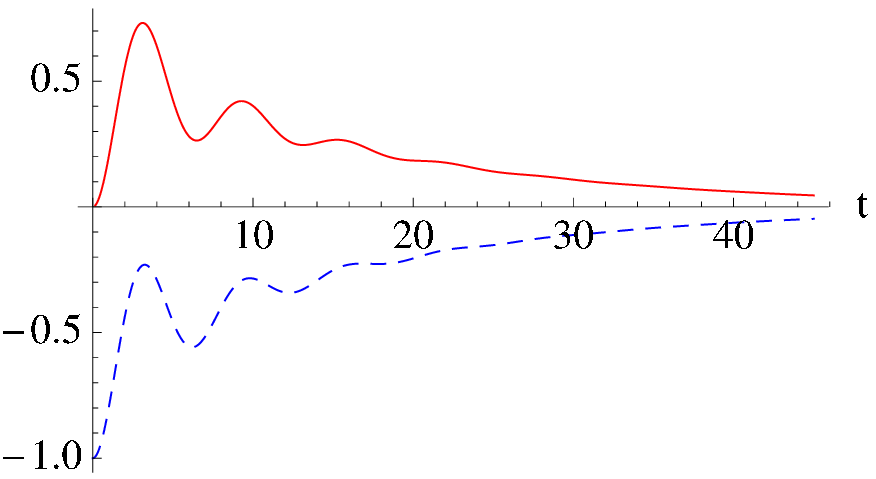}}\\
\subfigure[]{\includegraphics[width=5.2cm]{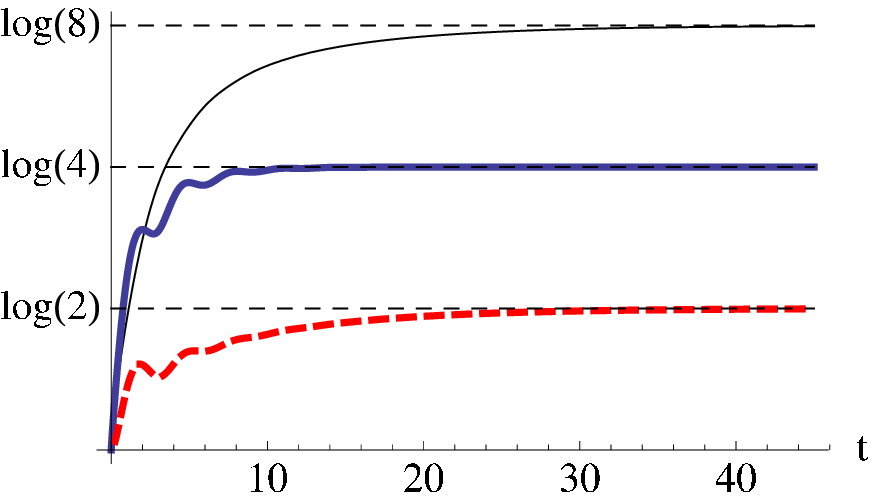}}
\subfigure[]{\includegraphics[width=5.1cm]{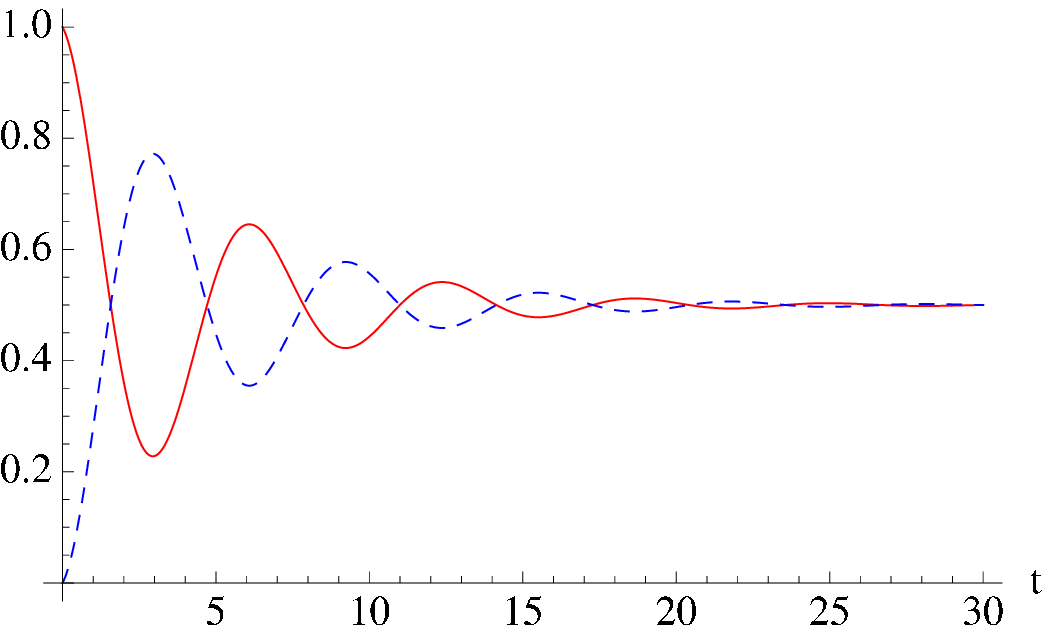}}
\caption{$\epsilon=0.1;\ \mu=0$; $\vartheta=\pi/2$.  (a) $\llrr{m_2(t),m_3(t)}\  0 \leq t \leq 90 \pi$; (b) as in figure 1. (c) Solid thin line: the entropy of the overall open system; solid thick line: the entropy of the cursor; dashed line: the entropy of the register. (d) Solid line: $Tr\llrrq{\rho_t\ (1+\tau_3(L))/2}$; dashed line:  $Tr\llrrq{\rho_t\ (1+\tau_3(L+1))/2}$.\label{fig:figFeynint}}
\end{figure}

\begin{figure}[h]
\centering
%
\subfigure[]{\includegraphics[width=4.7cm]{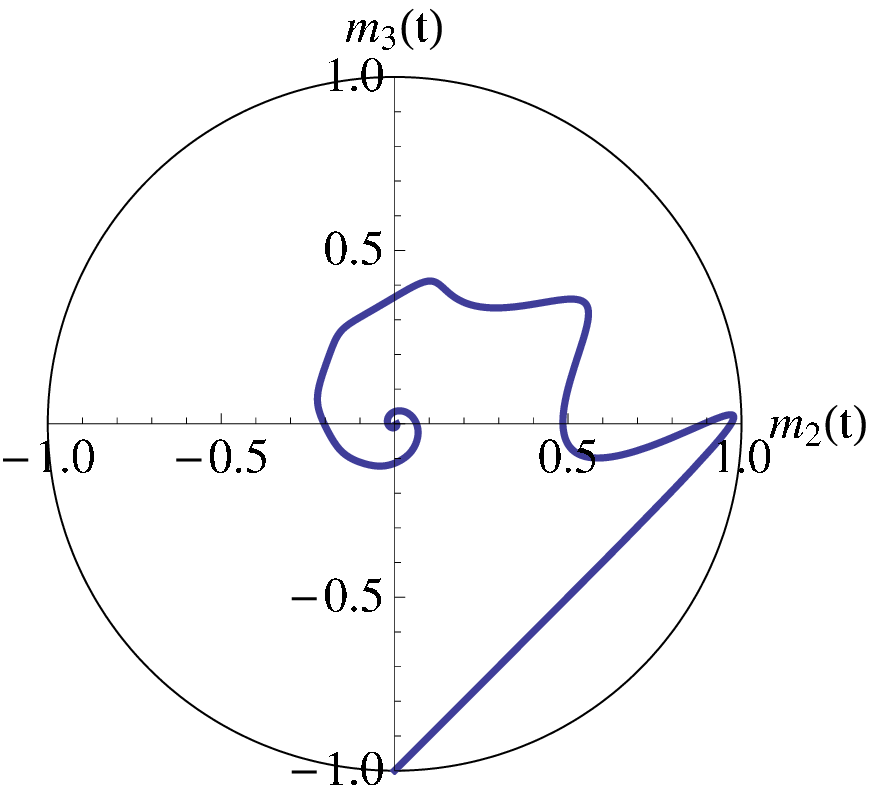}}
\subfigure[]{\includegraphics[width=5.2cm]{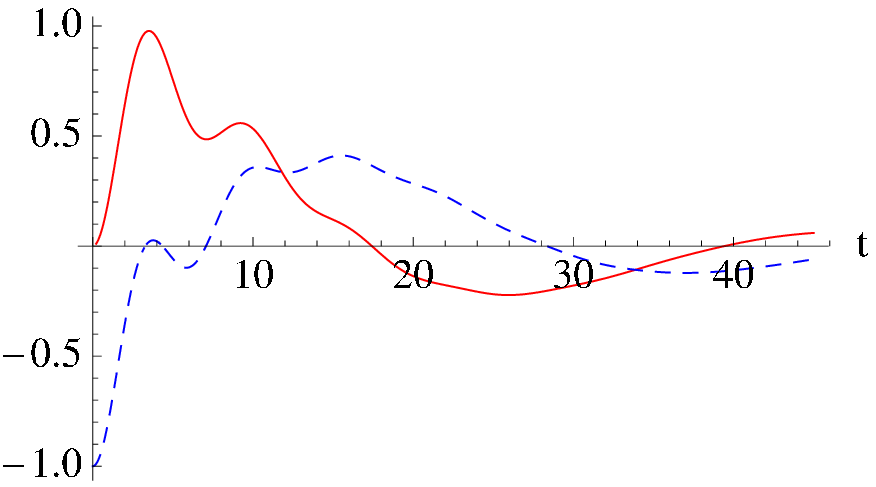}}\\
\subfigure[]{\includegraphics[width=5.2cm]{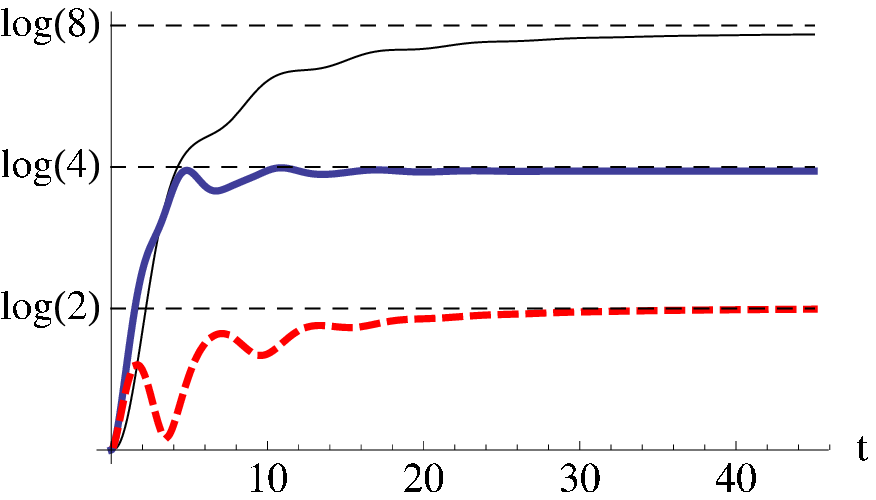}}
\subfigure[]{\includegraphics[width=5.02cm]{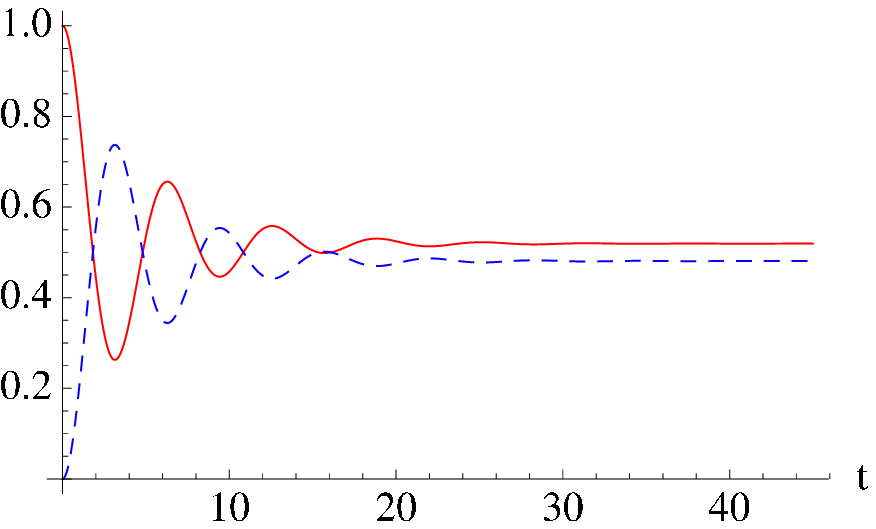}}
\caption{$\epsilon=0.1;\ \mu=-1$; $\vartheta=\pi/2$. Frames (a),(b),(c) and (d) as in \fref{fig:figFeynint}.\label{fig:figFeynintpot}}
\end{figure}

\begin{figure}
\centering
%
\subfigure[]{\includegraphics[width=4.7cm]{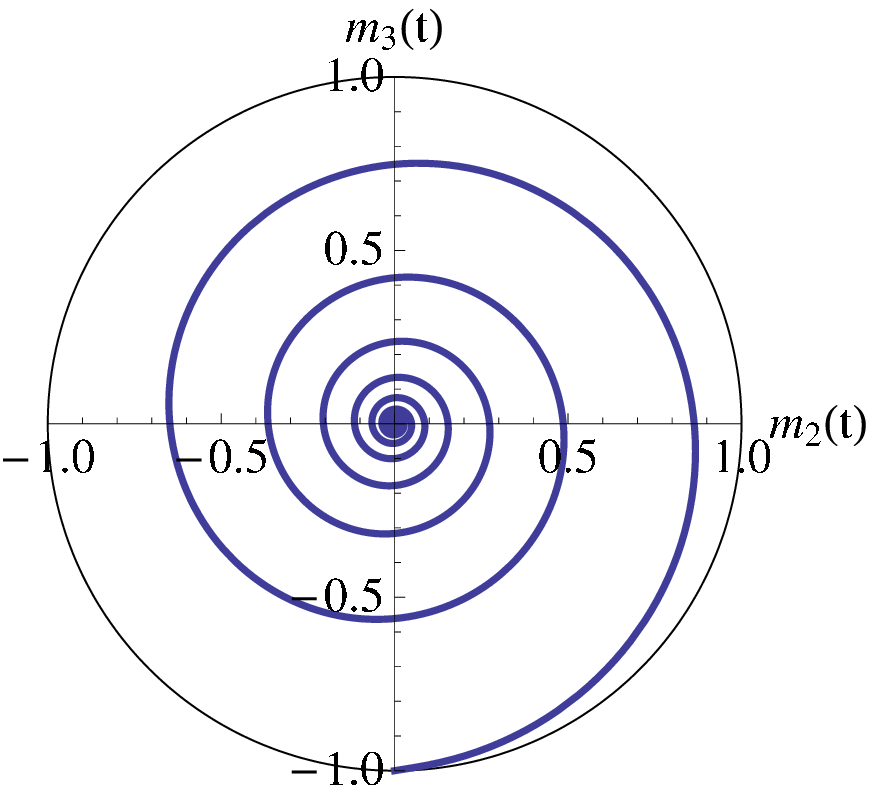}}
\subfigure[]{\includegraphics[width=5.2cm]{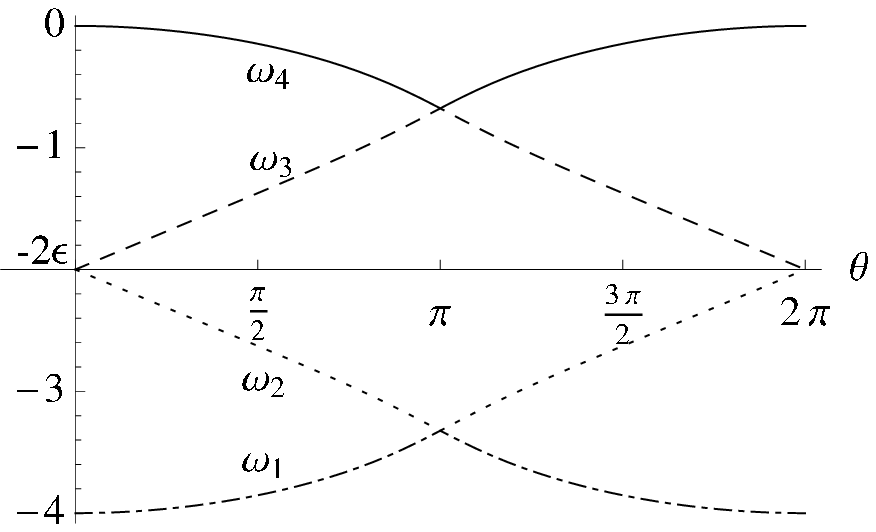}}\\
\subfigure[]{\includegraphics[width=5.2cm]{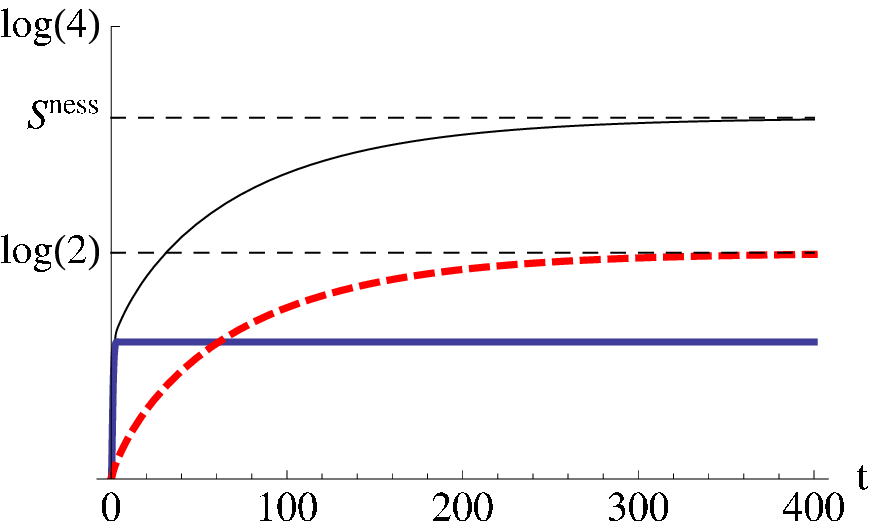}}
\subfigure[]{\includegraphics[width=5.2cm]{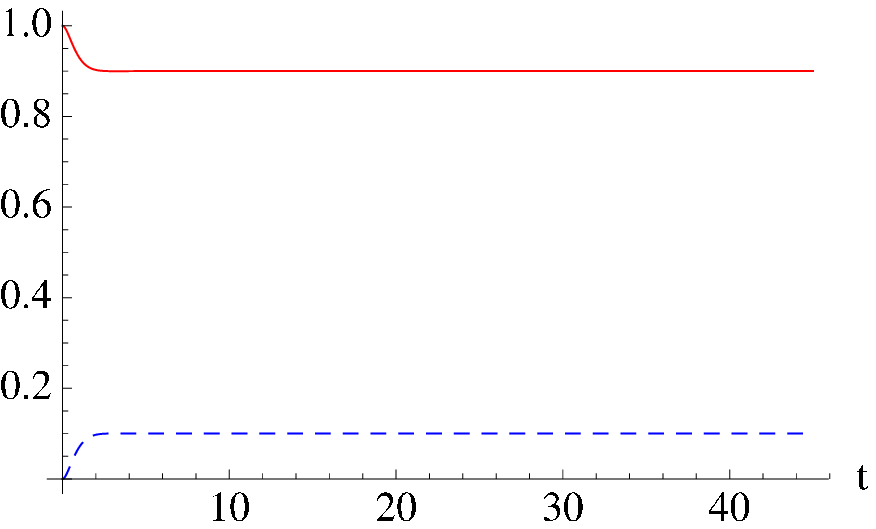}}
\caption{$\epsilon=1,\ \mu= -1, \vartheta= \pi/10.$ (a) Parametric plot of $(m_2(t),m_3(t))$. (b) $Re(\omega_1) < Re(\omega_2) <Re(\omega_3)< Re(\omega_4)$ as functions of $\vartheta$. (c) To compare with \ref{fig:figFeynint}(c) and \ref{fig:figFeynintpot}(c); this time the entropy of the register is larger than the entropy of the cursor. (d) To compare with \ref{fig:figFeynint}(d) and \ref{fig:figFeynintpot}(d).\label{fig:fignew}}
\end{figure}

Figures \ref{fig:figFeynint}(d) and \ref{fig:figFeynintpot}(d) show, for different values of the parameter $\mu$, the marginal distributions of $\tau_3(L)$ and $\tau_3(L+1)$ characterized, respectively, by the expectation values:
\begin{align*}
&Tr\llrrq{\rho_t \llrr{I+\tau_3(L)}/2}= \\
&= \frac{1+4 \epsilon^2 (1-\mu)}{2 \llrr{1+4 \epsilon^2}}+ e^{-2t \epsilon}\frac{\llrr{1+4 \epsilon^2 (1+\mu)}\cos(t) -2 \epsilon \mu \sin(t)}{2\llrr{1+4 \epsilon^2}},
\end{align*}
\begin{align*}
&Tr\llrrq{\rho_t \llrr{I+\tau_3(L+1)}/2}=\\
& = \frac{1+4 \epsilon^2 (1+\mu)}{2 \llrr{1+4 \epsilon^2}}- e^{-2t \epsilon}\frac{\llrr{1+4 \epsilon^2 (1+\mu)}\cos(t) -2 \epsilon \mu \sin(t)}{2\llrr{1+4 \epsilon^2}}.
\end{align*}
By comparison with \fref{fig:figFeynnoint}(d) it is immediate to notice, in both frames, the damping effect of the coupling parameter $\epsilon$ and, in \fref{fig:figFeynintpot}(d), the polarizing effect of the asymmetry parameter $\mu$: the left reservoir injects spins up in the chain while the right bath absorbs spins up, thus favoring a positive current. In this sense, when $\mu \neq 0$, the environment acts as a ``battery''\cite{landauer99}. 
\\A significant effect of the coupling with the environment can be seen in the entropy landscapes of figures \ref{fig:figFeynint}(c) and \ref{fig:figFeynintpot}(c). 
In both examples, the entropy of the overall system increases monotonically and the entropies of the two subsystems (register and cursor) differ form each other; the behaviour of the entropy of the cursor in the two cases is qualitatively similar.\\
What is interesting, in the short-time transient we are mainly interested in, is how differently the entropy of the register behaves in the absence or presence of a chemical potential difference between the reservoirs.\\
For $\mu=0$ (reservoirs at the same chemical potential) the entropy of the register increases almost monotonically, following the trend of the entropy of the cursor (\fref{fig:figFeynint}(c)).\\
For $\mu=-1$ (\fref{fig:figFeynintpot}(c)), around time $\pi$, the entropy of the register gets close to the value $0$, as it happened in the isolated system (\fref{fig:figFeynnoint}(c)). The presence of the ``battery''
is able to push, at $t=\pi$, the Bloch diagram closer to the boundary of the Bloch sphere (\fref{fig:figFeynintpot}(c)), which, we remind, corresponds to the register being in the desired output state with certainty.\\[5pt]
As the coupling with the environment increases, the polarization of $\tau(L)$ and $\tau(L+1)$ increases accordingly (\fref{fig:fignew}(d)). The presence of a ``strong'' battery results in the onset of a stationary current, witnessed by the uniform rotation of the register (see \fref{fig:fignew}(a)). The different relaxation times of the register and the cursor become evident, together with the dependence of the entropies of the cursor and of the overall system on the interaction parameters $\epsilon$, $\mu$ and $\vartheta$.
\\As a final example of this section, we exhibit in figure 6 the evolution of the mean chirality $Tr\llrr{\chi \ \rho_t}$ under Lindblad dynamics with the same parameters as in figure 4.
\begin{figure}[h]
\centering
\includegraphics[width=8cm]{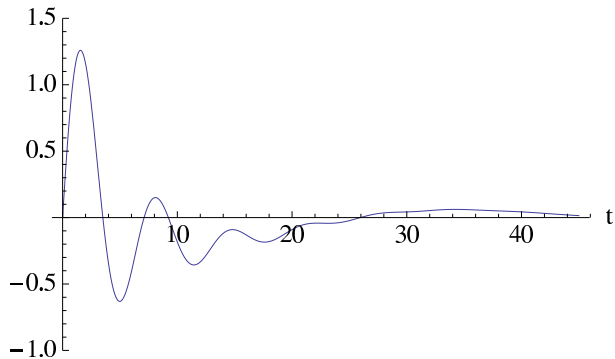}
\caption{$\epsilon=0.1;\ \mu=-1$; $\vartheta=\pi/2$. The mean chirality as a function of time. \label{fig:figchir}}
\end{figure}
%
%
\section{Conclusions and outlook \label{sec:conclusions}}
Our interest is in the effect of noise deteriorating the operation of the machine. We model by differences of chemical potential the action of a battery trying to compensate for the noise. The examples discussed in the previous section suggest that, as far as the short-time transient is concerned, the disruptive effects of the interaction with the environment can be mitigated by a proper tuning of the (macroscopic) reservoirs.\\
The toy model we discuss in this paper is an oversimplified version of the systems discussed in references 2 and 3. With respect to the two references just quoted, however, our simple model adds a peculiar difficulty: the links of the chain do not carry numerical hopping parameters, but additional quantum degrees of freedom, those of the register on which the computation is performed. This added ingredient sets the Lindblad equation \eref{eq:lindblad} out of the family of problems explicitly solvable by third quantization\cite{prosen08}.\\ 
In this note we used the entropy of the register to measure the ``degree of certainty'' (fidelity) with which the output state is reached; but \fref{fig:fignew} suggests also to use the register as a \emph{probe} to visualize what happens in the cursor when it interacts with the environment in more complicated ways. For example, one could investigate the onset of Anderson localization in a cursor with noisy hopping parameters.\\
An explicit solution of the model of \fref{fig:unit}, that is a system with an arbitrarily long cursor spin chain, would be most desirable. Our study of the structure of the master equation gives us some hint about how to tackle this general problem: as far as the dynamics of $System(+,+)$ and $System(-,-)$ is concerned, the solution scheme of third quantization \cite{prosen08} can be still applied. This means that observables of the cursor are already accessible. What requires a different approach is the solution of the subproblems $System(+,-)$ and $System(-,+)$, which contain all the information about the register and the register/cursor interaction.

\section*{Appendix: Spectrum of the Lindblad operator}
The analysis of the system \eref{eq:lindblad} of 64 equations is made simple by the fact that the commutation relations
\[
\llrrq{\sigma_1,H}=0; \ \llrrq{\sigma_1,L_j} = 0,\ j=1,\ldots,4
\]
and the adoption of the basis $\left \{\ket{e_k},\ k=1,\ldots,8 \right \}$ make it possible to split the problem into 4 uncoupled systems of 16 equations. Setting, in what follows, $\rho_t(j,k)=\bra{e_j}\rho_t \ket{e_k}$, the four systems can be described in the following way:
\\[5pt]\emph{System}$(+,+)$ relates matrix elements of $\rho_t$ between vectors of the basis $\{\ket{e_k}\}$ both belonging to the eigenvalue $+1$ of $\sigma_1$ and is of the form:
\[
\frac{d\rho_t(j,k)}{dt} = \sum_{m=1,\ldots,4 \atop n=1,\ldots,4} M_{j,k;m,n}^{++}\rho_t(m,n),\  1 \leq j,k \leq 4.
\]
The $16 \times 16$ matrix $M^{++}$ turns out to have characteristic polynomial:
\begin{align*}
&\det(M^{++}-x I) = x \llrr{1+ (x+2\epsilon)^2 }\llrr{x+4\epsilon}\llrr{x+2 \epsilon}^4 \\
&\llrr{x^2 +4 \epsilon x +3 \epsilon^2 - \epsilon \mu +\frac{1}{4}}^2 \llrr{x^2 +4 \epsilon x +3 \epsilon^2+ \epsilon \mu +\frac{1}{4}}^2.
\end{align*}
Under the initial condition \eref{eq:initialcond} only the following six components of the solution of $System(+,+)$ do not vanish identically:
\begin{align*}
\rho_t(1,1) &= \frac{1+4 \epsilon^2(1-\mu^2)}{8(1+4 \epsilon^2)}+ e^{t\llrr{i-2\epsilon}} \frac{\epsilon \mu \llrr{2 \epsilon(1+\mu)-i}}{4(1+4 \epsilon^2)} + \\
& +e^{t\llrr{-i-2\epsilon}} \frac{\epsilon \mu \llrr{2 \epsilon(1+\mu)+i}}{4(1+4 \epsilon^2)} -e^{t(-4\epsilon)}\frac{1+4 \epsilon^2(1+\mu)^2}{8(1+4 \epsilon^2)}; \\
\rho_t(2,2) &= \frac{1+4 \epsilon^2(1-\mu)^2}{8(1+4 \epsilon^2)}+ e^{t\llrr{i-2\epsilon}} \frac{1+4 i \epsilon \mu +4 \epsilon^2 \llrr{1-\mu^2}}{8(1+4 \epsilon^2)}  + \\
& +e^{t\llrr{-i-2\epsilon}} \frac{1-4 i \epsilon \mu +4 \epsilon^2 \llrr{1-\mu^2}}{8(1+4 \epsilon^2)}+ e^{t(-4\epsilon)}\frac{1+4 \epsilon^2(1+\mu)^2}{8(1+4 \epsilon^2)};\\
\rho_t(3,3) &=\frac{1+4 \epsilon^2(1+\mu)^2}{8(1+4 \epsilon^2)}\llrr{1-e^{t(i-2\epsilon)}-e^{t(-i-2\epsilon)}+e^{t(-4\epsilon)}};\\
\rho_t(4,4)&=\frac{1}{2} - \llrr{\rho_t(1,1)+\rho_t(2,2)+\rho_t(3,3)};\\
\rho_t(2,3)&= \overline{\rho_t(3,2)} = \\
& = e^{i \vartheta/2} \llrr{\frac{i \epsilon \mu}{2\llrr{1+4 \epsilon^2}} +  e^{t(i-2 \epsilon)}\frac{i - 2 \epsilon (1+\mu)}{8(2\epsilon-i)} +e^{t(-i-2 \epsilon)}\frac{i + 2 \epsilon (1+\mu)}{8(2\epsilon+i)}  }.
\end{align*}
%
\emph{System}$(-,-)$ relates matrix elements of $\rho_t$ between vectors of the basis $\{\ket{e_k}\}$ both belonging to the eigenvalue $-1$ of $\sigma_1$; it is of the form:
\[
\frac{d\rho_t(j,k)}{dt} = \sum_{m=5,\ldots,8 \atop n=5,\ldots,8} M_{j,k;m,n}^{--}\rho_t(m,n),\  5 \leq j,k \leq 8.
\]
The $16 \times 16$ matrix $M^{--}$ can be obtained from $M^{++}$ by changing $\vartheta$ into $-\vartheta$ and has, therefore, the same eigenvalues. By the same argument it is easy to check that under \eref{eq:initialcond} the only non vanishing components of the solution of $System(-,-)$ are:
\begin{align*}
&\rho_t(5,5) = \rho_t(1,1),\ \rho_t(6,6) = \rho_t(2,2),\\
&\rho_t(7,7) = \rho_t(3,3),\ \rho_t(8,8)=  \rho_t(4,4),\\
&\rho_t(6,7)  = e^{-i\vartheta} \rho_t(2,3),\ \rho_t(7,6) = e^{i \vartheta} \rho_t(3,2).
\end{align*}
\emph{System}$(+,-)$ relates matrix elements $\bra{e_j} \rho_t \ket{e_k}$, where \ket{e_j} belongs to the eigenvalues $+1$ of $\sigma_1$ and \ket{e_k} belongs to the eigenvalues $-1$ of $\sigma_1$.
\begin{align*}
&\frac{d\rho_t(j,k)}{dt} = \sum_{m=1,\ldots,4 \atop n=5,\ldots,8}  M_{j,k;m,n}^{+-}\rho_t(m,n),\  1 \leq j \leq 4,\ 5 \leq k \leq 8.\\
&\det(M^{+-}-xI) = \frac{1}{2^8} \llrr{x+2\epsilon}^4 \llrr{x-\omega_1}\llrr{x-\omega_2}\llrr{x-\omega_3}\llrr{x-\omega_4}\\
& [1+8\llrr{ 2x^4+16 \epsilon x^3 +\llrr{1+44 \epsilon^2} x^2 +4\llrr{\epsilon+12\epsilon^3}x+\epsilon^2\llrr{4+18 \epsilon^2-\mu^2} } +\\
&-8\epsilon^2 \llrr{ \llrr{1+\mu^2}\cos(\vartheta)+2i\mu\sin(\vartheta)  } ]^2.
\end{align*}
Here $\llrr{\omega_1,\omega_2,\omega_3,\omega_4}$ is any ordering (for instance by increasing real part) of the four eigenvalues
\begin{align*}
&\omega_{\pm,\pm} = -2 \epsilon \pm\\
& \pm \frac{1}{\sqrt{2}} \sqrt{4 \epsilon^2 -1 
\pm 
\sqrt{1+16\epsilon^4 -8 \epsilon^2 \mu^2 + 8 \epsilon^2 \llrr{\cos\llrr{\vartheta}+\mu^2 \cos(\vartheta)+2i\mu \sin(\vartheta) }}}
\end{align*}
of $M^{+-}$ that actually contribute to the evolution of the non vanishing components of the solution under the initial condition \eref{eq:initialcond}. 

In terms of the eigenvalues $\omega_{\pm,\pm}$, we can write:
\begin{align*}
&\rho_t(1,5) = \rho_t(4,8) = \\
&= \frac{\epsilon}{4} \sum_{j=1}^4 e^{\omega_j t} \frac{(\mu-1) e^{-i\vartheta}+(\mu+1)\llrr{-1+8\epsilon^2(\mu-1)+4 \epsilon(\mu-2) \omega_j -2\omega_j^2}}{\prod_{1 \leq k \leq 4 \atop k \neq j} \llrr{\omega_j - \omega_k}}\\
&\rho_t(2,6) =- \frac{1}{4} \sum_{j=1}^4 e^{\omega_j t} \frac{(2 \epsilon + \omega_j)\llrr{1+4\epsilon^2(\mu-1)^2-4 \epsilon(\mu-2) \omega_j +2\omega_j^2}}{\prod_{1 \leq k \leq 4 \atop k \neq j} \llrr{\omega_j - \omega_k}},\\
&\rho_t(3,7) =- \frac{e^{-i \vartheta} \llrr{1+ 4 e^{i \vartheta} \epsilon^2(1+\mu)^2}}{4} \sum_{j=1}^4 e^{\omega_j t} \frac{(2 \epsilon + \omega_j)}{\prod_{1 \leq k \leq 4\atop  k \neq j} \llrr{\omega_j - \omega_k}},\\
&\rho_t(2,7)=-\rho_t(3,6) =\\
&= \frac{ie^{-i \vartheta/2}}{4} \sum_{j=1}^4 e^{\omega_j t} \frac{-2 \epsilon^2\llrr{e^{i \vartheta}(\mu+1)^2 -(\mu-1)^2 } - 2 \epsilon(\mu-2)\omega_j + \omega_j^2}{\prod_{1 \leq k \leq 4 \atop k \neq j} \llrr{\omega_j - \omega_k}}.
\end{align*}
%
We notice, for completeness, that the spectrum of $M^{+-}$ includes also the following eigenvalues (which do not contribute to evolution under the initial condition \eref{eq:initialcond}), each with multiplicity 2:
\begin{align*}
& \zeta_{\pm,\pm} = 
-2 \epsilon \pm \frac{e^{-i \vartheta}}{2} \sqrt{e^{2i\vartheta}\llrr{4 \epsilon^2 - 1}\pm 2 e^{i \vartheta} \sqrt{e^{i \vartheta} \epsilon^2 \llrr{\mu-1 + e^{i \vartheta}\llrr{\mu+1}  }^2      }  }.       
\end{align*}
\emph{System}$(-,+)$: relates matrix elements $\bra{e_j} \rho_t \ket{e_k}$, where \ket{e_j} belongs to the eigenvalues $-1$ of $\sigma_1$ and \ket{e_k} belongs to the eigenvalues $+1$ of $\sigma_1$. It is of the form:
\[
\frac{d\rho_t(j,k)}{dt} = \sum_{m=5,\ldots,8 \atop n=1,\ldots,4}  M_{j,k;m,n}^{-+}\rho_t(m,n),
\] for $5 \leq j \leq 8,\ 1 \leq k \leq 4$.\\
The $16 \times 16$ matrix $M^{-+}$ can be obtained from $M^{+-}$ by Hermitian conjugation. The solution of $System(-,+)$ is, obviously, obtained from the solution of $System(+,-)$ by Hermitian conjugation.\\[5pt]
We observe that, as only $M^{++}$ and $M^{--}$ have the eigenvalue 0, the initial condition $\rho_0$ converges, as $t \to \infty$, to a Nonequilibrium Stationary State $\rho_{ness}$ represented by the density matrix having the following non vanishing matrix elements:
\begin{align*}
&\rho_{ness}(1,1) = \rho_{ness}(4,4)=\rho_{ness}(5,5)=\rho_{ness}(8,8) = \frac{1+4 \epsilon^2 \llrr{1-\mu^2}}{8 \llrr{1+4 \epsilon^2}}\\
&\rho_{ness}(2,2) = \rho_{ness}(6,6) = \frac{1+4 \epsilon^2 \llrr{1-\mu}^2}{8 \llrr{1+4 \epsilon^2}}\\
&\rho_{ness}(3,3) = \rho_{ness}(7,7) = \frac{1+4 \epsilon^2 \llrr{1+\mu}^2}{8 \llrr{1+4 \epsilon^2}}\\
&\rho_{ness}(2,3) = \overline{\rho_{ness}(3,2)} =-\rho_{ness}(7,6) = -\overline{\rho_{ness}(6,7)} =  \frac{i \epsilon \mu e^{i \vartheta/2}}{2 \llrr{1+ 4 \epsilon^2}}.
\end{align*}
It is immediate to check that this state has entropy:
\begin{align*}
&S^{ness} = \log(8) - \log\llrr{1-\frac{4 \epsilon^2 \mu^2}{1+4 \epsilon^2}}- \frac{2 \epsilon \mu \tanh^{-1} \llrr{\frac{4 \epsilon \mu \sqrt{1+4 \epsilon^2}}{1+4 \epsilon ^2 \llrr{1+\mu^2}}}}{\sqrt{1+4 \epsilon^2}}.
\end{align*}



\begin{thebibliography}{10}

\bibitem{bose03}
S.~Bose.
\newblock Quantum communication through an unmodulated spin chain.
\newblock {\em Phys. Rev. Lett.}, 91(20):207901, 2003.

\bibitem{dhar03}
A.~Dhar and B.~Shastry.
\newblock Quantum transport using the {F}ord-{K}ac-{M}azur formalism.
\newblock {\em Phys. Rev. B}, 67(19):195405, 2003.

\bibitem{roy07}
D.~Roy and A.~Dhar.
\newblock Electron transport in a one dimensional conductor with inelastic
  scattering by self-consistent reservoirs.
\newblock {\em Phys. Rev. B}, 75(19):195110, 2007.

\bibitem{wichterich07}
H.~Wichterich, M.~J. Henrich, H.-P. Breuer, J.~Gemmer, and M.~Michel.
\newblock Modeling heat transport through completely positive maps.
\newblock {\em Physical Review E}, 76(3):031115, 2007.

\bibitem{znidaric10}
M.~\v{Z}nidaric.
\newblock Quantum transport in 1d systems via a master equation approach:
  numerics and an exact solution.
\newblock {\em arXiv:1012.4684v1}.

\bibitem{breuer04}
H.-P. Breuer, D.~Burgarth, and F.~Petruccione.
\newblock Non-markovian dynamics in a spin star system: Exact solution and
  approximation techniques.
\newblock {\em Phys. Rev. B}, 70(4):045323, 2004.

\bibitem{prosen08}
T.~Prosen.
\newblock Third quantization: a general method to solve master equations for
  quadratic open {F}ermi systems.
\newblock {\em New J. Phys.}, 10:043026, 2008.

\bibitem{wilson74}
K.~G. Wilson.
\newblock Confinement of quarks.
\newblock {\em Physical Review D}, 10(8):2445--2459, 1974.

\bibitem{feyn86}
R.P. Feynman.
\newblock Quantum mechanical computers.
\newblock {\em Found. Phys.}, 16(6):507--31, 1986.

\bibitem{landauer99}
Y.~Imry and R.~Landauer.
\newblock Conductance viewed as transmission.
\newblock {\em Rev. Mod. Phys.}, 71(2):S306--S312, 1999.

\bibitem{peres85}
A.~Peres.
\newblock Reversible logic and quantum computers.
\newblock {\em Phys. Rev. A}, 32(6):3266--3276, 1985.

\bibitem{cirac08}
K.~G.~H. Vollbrecht and J.~I. Cirac.
\newblock Quantum simulators, continuous-time automata, and translationally
  invariant systems.
\newblock {\em Phys. Rev. Lett.}, 100(1):010501, 2008.

\bibitem{defa06b}
D.~de~Falco and D.~Tamascelli.
\newblock Entropy generation in a model of reversible computation.
\newblock {\em RAIRO: Inf. Theor. Appl.}, 40:93--105, 2006.

\bibitem{nagaj10}
D.~Nagaj.
\newblock Fast universal quantum computation with railroad-switch local {H}amiltonians.
\newblock {\em J. Math. Phys.}, 51:062201, 2010.

\bibitem{defa06a}
D.~de~Falco and D.~Tamascelli.
\newblock Speed and entropy of an interacting continuous time quantum walk.
\newblock {\em J. Phys. A: Math. Gen.}, 39:5873--5895, 2006.

\bibitem{tsomokos08}
D.~I.~Tsomokos, J.~J.~ Garc\'ia-Ripoll, N.~R.~Cooper and J.~K.~Pachos.
\newblock Chiral entanglement in triangular lattice models.
\newblock {\em Phys. Rev. A}, 77(1):012106, 2008.

\bibitem{benenti09}
G.~Benenti, G.~Casati, T.~Prosen, D.~Rossini and M.~\ifmmode \check{Z}\else \v{Z}\fi{}nidari\ifmmode \check{c}\else \v{c}\fi{}.
\newblock Charge and spin transport in strongly correlated one-dimensional quantum systems driven far from equilibrium.
\newblock {\em Phys. Rev. B}, 80(3):035110, 2009.

\end{thebibliography}
%
\end{document}